\documentclass[aps,prb,showpacs,floatfix,twocolumn,superscriptaddress]{revtex4-1}
\usepackage{amssymb}
\usepackage{graphicx}
\usepackage{color}
\usepackage{amsmath}
\usepackage{amsbsy}
\usepackage{url}
\usepackage{dsfont}
\usepackage[colorlinks=true,dvipdfm]{hyperref}
\hypersetup{
     linktocpage=true,
     hyperindex=true,
     bookmarks=true,
     bookmarksopen=true,
     bookmarksnumbered=true,
     allcolors=blue,
     pdfborder=0 0 0,
}

\begin{document}

\title{Stationary Josephson effect in a short multi-terminal junction}

\date{\today}

\author{Sebastian~Mai}
\affiliation{Theoretische Physik III, Ruhr-Universit\"{a}t Bochum, D-44780 Bochum, Germany}
\affiliation{IPhT, CEA-Saclay, L'Orme des Merisiers, 91191 Gif-sur-Yvette, France}
\author{Ervand~Kandelaki}
\affiliation{Theoretische Physik III, Ruhr-Universit\"{a}t Bochum, D-44780 Bochum, Germany}
\affiliation{IPhT, CEA-Saclay, L'Orme des Merisiers, 91191 Gif-sur-Yvette, France}
\author{Anatoly~Volkov}
\affiliation{Theoretische Physik III, Ruhr-Universit\"{a}t Bochum, D-44780 Bochum, Germany}
\author{Konstantin~Efetov}
\affiliation{Theoretische Physik III, Ruhr-Universit\"{a}t Bochum, D-44780 Bochum, Germany}
\affiliation{IPhT, CEA-Saclay, L'Orme des Merisiers, 91191 Gif-sur-Yvette, France}
\begin{abstract}
We study the dc Josephson effect and the density of states in a multiterminal structure of cross-type geometry that consists of four superconducting  electrodes connected by one-dimensional normal or ferromagnetic wires. We find that the Josephson current $I_{Jz}$ has a sinusoidal dependence on the phase difference $\phi_z$ between the superconductors in the horizontal wire
with a critical current $I_c$ that can be varied by changing the phase difference $\phi_y$ between the superconductors in the vertical wire. The period of the function $I_{Jz}(\phi_z)$ depends on the ratio of the interface resistances $R_{Sn,z}/R_{Sn,y}$ being equal to $2\pi $ if this ratio is small and equal to $4\pi $ in the opposite limit. We also calculate the amplitudes of both the singlet and the odd-frequency triplet components in the system  under consideration. The triplet component amplitude may be significantly larger than the amplitude of the singlet component.
\end{abstract}

\pacs{74.45.+c, 74.50.+r, 74.78.Fk}
\maketitle

\section{Introduction}
\label{sec-Intro}

In the recent years, there has been an increasing interest in studies of Josephson junctions (JJs) with controllable parameters. The value and sign
of the Josephson critical current $I_c$ can be changed by variation of
temperature and external or internal parameters. To some extent, this
interest is caused by the possibility of technological applications of such
JJs, for example in Q bits (for a review see, e.g., Ref.~\onlinecite{Makhlin}).

A remarkable example of a JJ with properties that depend on internal parameters in a crucial way is a superconductor/ferromagnet/superconductor
(S/F/S) junction. A specific action of the exchange field on the spins of Cooper pairs leads, in a certain range of temperatures and thicknesses of the ferromagnetic films, to a sign change of the critical current $I_c$ in S/F/S JJs \cite{BuzdinRMP}. This effect predicted long time ago  \cite{Bulaev77,BuzdinBul} has been observed by many experimental groups
\cite{Ryaz1,Ryaz2,Kontos,Blum,Bauer,Sellier,Blamire06,Weides06,Weides2,Westerholt09}.
In S/F/S JJs consisting of ordinary singlet $s$-wave superconductors and a
ferromagnet with a nonhomogeneous magnetization not only a singlet component
arises in the F due to proximity effect, but also a nontrivial, so-called
odd triplet component that penetrates the ferromagnetic layer over a long
distance (for a review see  Refs.~\onlinecite{BVErmp,EschrigPhysToday,Efetov12}). This component
observed in many experiments  \cite{Klapwijk,Sosnin,Birge10,Robinson,Westerholt10,Aarts,Birge12}
may also lead to an unusual and controllable behavior of JJs.

The latter property has been demonstrated in recent experiments \cite{Birge12} on a multi-layered S/F$'$FFF$'$/S Josephson junction in which the F$'$ and F layers are a weak (F$'$ $\rightarrow \mathrm{Ni}$ or $\mathrm{Pd}_{0.88}\mathrm{Ni}_{0.12}$) and a strong (F$\rightarrow \mathrm{Co}$) ferromagnet, respectively. Since the singlet component leaking
from the superconductor S (S $\rightarrow \mathrm{Nb}$) cannot penetrate the rather thick Co layer ($d_{\mathrm{Co}}=20\mathrm{nm}$), the Josephson coupling is provided by the odd triplet component arising due to noncollinearity of magnetization vectors $\mathbf{M}$ in the F$'$ and F layers. The maximum amplitude of this triplet component occurs at perpendicular orientation \cite{Buzdin07,VE10,Radovic10}. Applying an external magnetic field, the authors of Ref.~\onlinecite{Birge12}
changed the mutual orientation of the magnetization vectors in the F$'$ and F layers and could increase the critical current $I_c$ more than by an order of magnitude.

In S/FF/S JJs where the magnetization vectors $\mathbf{M}$ in the F layers are parallel or antiparallel with respect to each other, the critical current $I_c$ depends on the mutual orientations of the $\mathbf{M}$ vectors (see theoretical papers Refs.~\onlinecite{BVE01,Blanter,Zaitsev09}). This dependence was measured by Robinson \emph{et al.}\cite{BlamirePRL} In accordance with the theoretical predictions, they found that the current $I_c$ is larger for the antiparallel orientation than for the parallel one.

The change of sign of the critical Josephson current may be realized not
only in S/F/S JJs, but also in superconductor/normal metal/superconductor (S/N/S) JJs (see theoretical papers Refs.~\onlinecite{Volkov95,Yip,Wilhelm}) if the quasiparticle distribution function in the normal metal is not in equilibrium with the superconducting reservoirs. This disequilibrium may be achieved, e.g., in a four terminal JJ by passing a dissipative current trough the lateral N reservoirs. Then, the critical current $I_c(V)$ decreases as a function of the voltage $V$ applied between the N reservoirs and may become negative. This effect has been established in experiments by Baselmans \emph{et al.}\cite{KlapwijkWees}

The critical current $I_c$ in multi-terminal JJs may not only decrease
under non-equilibrium conditions, but may also increase \cite{Seviour2000}. A weak increase of the Josephson current has been observed in Josephson weak
links under microwave radiation several decades ago  \cite{Dayem,Latyshev,Klapwijk76,Klapwijk77,Wolter}. The observed increase of $I_c$ is probably related to the stimulation of superconductivity by microwave irradiation predicted by Eliashberg \cite{Eliashberg}.

Multi-terminal S/N/S JJs were studied in several papers  \cite{KlapwijkWees,Petrashov93,Petrashov00}. These junctions contain a dissipative element, a normal conductor or leads through which a dissipative current was passed. For example, in experiments by Petrashov \emph{et al.},
\cite{Petrashov93,Petrashov00,PetrashovCMMP11} the modulation of the conductance between the N reservoirs due to a variation of the phase difference $\varphi$ between two superconductors in four-terminal structures (such structures are called Andreev interferometers) has been measured.

The possibilities to change the critical Josephson current in non-dissipative four-terminal structures by passing current through lateral electrodes were studied theoretically in Refs.~\onlinecite{Omelyan01,Omelyan02} (ballistic limit) and in a recent Ref.~\onlinecite{Linder12} (dirty case). The authors of Ref.~\onlinecite{Linder12} assumed a weak proximity effect. Physically, this means that the interface resistance is much larger than the resistance of the normal two-dimensional region between four superconducting leads. In this limit, the Usadel equation can be linearized. Solving this linearized equation for a rectangular geometry, Alidoust \emph{et al.} \cite{Linder12} have found the dependence of the critical current $I_c$ between the left and right superconductors on the phase difference between the two other electrodes (upper and down).

In this paper, we consider a four-terminal Josephson junction of cross-type geometry as shown in Fig.~\ref{fig-setup}. It consists of a normal metal (n) wire or film, which connects four superconducting reservoirs. In our analysis, the connecting wires of the system are considered as one dimensional. The normal metal wires have the total lengthes $2L_{\nu}$ $(\nu=z,y)$ along the corresponding axis $\nu$. Unlike Ref.~\onlinecite{Linder12}, we analyze the
case of arbitrary strength of proximity effect. Using the Usadel equation and assuming that the Thouless energy $E_{\mathrm{Th}} = D/L^2$ is large enough [$E_{\mathrm{Th}} \gg \mathrm{max}(T,\Delta)$], we find an exact solution for the quasiclassical Green's functions. Here, $D=v_Fl/3$ is the diffusion coefficient, $v_F$ is the Fermi velocity, $l$ is the electronic mean free path, $T$ denotes temperature, $\Delta$ is the superconducting gap parameter, $L=L_z+L_y$, and we choose units such that $\hbar = k_B \equiv 1$. Note that the case of a high Thouless energy compared to the exchange field was analysed in Refs.~\onlinecite{Fominov,Nazarov} for another SF system. We consider temperatures $T$ far below the critical temperature $T_c$ of the
superconducting electrodes so that the temperature dependence of the gap function $\Delta(T)$ can be neglected, i.e., $\Delta$ represents the $T=0$ value of $\Delta(T)$. With the help of these functions we calculate the density of states (DOS) at the crossing point and find the dependence of the critical Josephson current in the vertical JJ on the phase difference in the horizontal JJ. In the considered ``zero-dimensional" limit, expressions for the quasiclassical Green's functions are more compact and describe the considered JJ for any ratio of the interface resistance to the resistance of the n wires. In particular, they describe gapped states of the n wires and the dependence of the induced gap on the phase difference.

We also study the case of ferromagnetic wires connecting the superconductors (see Sec.~\ref{sec-Fwire}). It is assumed that the magnetization $\mathbf{M}$ in the wires is oriented along the wires and that the ferromagnets are not strong so that the condensate penetrates into the ferromagnets and the condition $E_{\mathrm{Th}}\gg V_{\mathrm{ex}}$ is fulfilled, where $V_{\mathrm{ex}}$ denotes the exchange energy. We show that not only the singlet component arises in the ferromagnetic cross, but also the odd-frequency $s$-wave triplet component with non-zero projection of the total spin $S$ on the $\mathbf{M}$ vectors. The amplitude of this triplet component may be significantly larger than the amplitude of the singlet component. In the concluding section, we discuss the obtained results.

\section{Model and Basic Equations}
\label{sec-model}
\begin{figure}[t]
\includegraphics{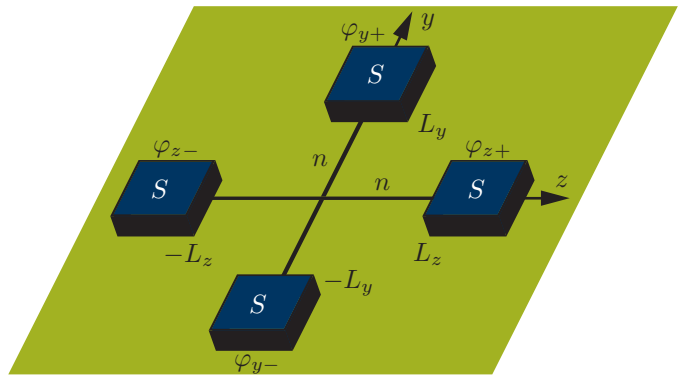}
\caption{(Color online) Schematic setup of a four-terminal Josephson junction  of cross-type geometry consisting of four superconducting reservoirs connected by one-dimensional normal metal wires. The junction is assumed to be located in the $zy$ plane with interfaces at $\nu=\zeta L_{\nu}$, where $\nu=z,y$ and $\zeta=\pm$. The superconductors are supposed to be identical but with different phases $\varphi_{\nu\zeta}$ of the order parameter.}
\label{fig-setup}
\end{figure}

We consider an S/S structure of cross-type geometry as shown in  Fig.~\ref{fig-setup}. In the horizontal and vertical directions, we have a $\mathrm{S}(-L_{z})-\mathrm{n}\big/\mathrm{n}-\mathrm{S}(L_{z})$ and
$\mathrm{S}(-L_{y})-\mathrm{n}\big/\mathrm{n}-\mathrm{S}(L_{y})$
structure, respectively. We will employ the method of quasiclassical Green's functions, which is well developed and can be used in problems where quantum effects on the scale of the Fermi wave length are unimportant \cite{Rammer,BelzigRev,Kopnin}. In particular, this theory was applied to the study of multiterminal S/S and S/N circuits in many papers (see, e.g.,
Refs.~\onlinecite{VZK93,Nazarov94,Volkov95,Zaitsev99,Kandelaki10,Linder12} and references therein).

We study the case of dirty superconductors that corresponds to most experimental situations, i.e., we assume that the electronic mean free path $l$ is much shorter than the superconducting coherence length $\xi_S=\sqrt{D/\Delta}$ and the lengthes of the n wires, $L_{\nu}$.
In the considered equilibrium case, one needs to find only the retarded (advanced) Green's function in order to obtain thermodynamical quantities as the density of states or the Josephson current. For singlet superconductors as assumed here, these functions have a $2\times 2$ matrix structure
in the particle-hole space. We denote the quasiclassical temperature Green's function by $\hat{g}(\omega)$ where $\omega = (2n+1)\pi T$ is the Matsubara frequency. In the n wires, the matrix function $\hat{g}$ obeys the Usadel equation \cite{Usadel}

\begin{equation}
- \partial_{\nu} \left( \hat{g} \cdot \partial _{\nu}\hat{g} \right)
+ \kappa_{\omega}^2 \left[ \hat{\tau}_3, \hat{g} \right]
=
0
\label{eq-UsadelEqn}
\end{equation}
and the normalization condition

\begin{equation}
\hat{g} \cdot \hat{g} = \hat{\mathds{1}},
\label{eq-Norm}
\end{equation}
where $\kappa_{\omega}^2=\omega/D$, $\hat{\tau}_{3}$ is the third Pauli matrix, and $\hat{\mathds{1}}$ is the $2\times2$ unity matrix in the particle-hole space. The boundary conditions have the form \cite{K+L}

\begin{equation}
\hat{g} \cdot \partial_{\nu}\hat{g}\big|_{\nu=\zeta L_{\nu}}
=
\zeta \kappa_{\nu} \left[ \hat{g},\hat{g}_S \right]\big|_{\nu=\zeta L_{\nu}},
\label{eq-BoundaryConds}
\end{equation}
where $\zeta=\pm$, $\kappa_{\nu}=\rho /2R_{nS,\nu}$, $\rho$ is the specific resistance of the n wire, $R_{nS,\nu}$ is the nS interface resistance at $\nu=\zeta L_{\nu}$ per unit area, and we assume that  $R_{nS,\nu}(L_{\nu})=R_{nS,\nu}(-L_{\nu})$.

As was noted above, the condition

\begin{equation}
L_{\nu}^2 \ll \mathrm{min}\{D/T, D/\Delta\}
\label{eq-ShortJunctionCond}
\end{equation}
is supposed to be fulfilled. Therefore one can integrate  Eq.~(\ref{eq-UsadelEqn}) over $\nu$ from $-L_{\nu}$ to $0$ and from $0$ to $L_{\nu}$ regarding the Green's function $\hat{g}$ as nearly constant. Taking into account the boundary condition (\ref{eq-BoundaryConds}), we obtain the equations

\begin{equation}
\zeta \hat{g} \cdot \partial_{\nu}\hat{g} \big|_{\nu=\zeta 0} - \kappa_{\nu} \left[ \hat{g},
\hat{g}_S(\zeta L_{\nu}) \right]
=
- \kappa_{\omega}^2 L_{\nu} \left[\hat{\tau}_3, \hat{g} \right].
\label{eq-BoundaryCondsIntegrated}
\end{equation}
Then, we sum up all these equations and take into account the law of  ``current" conservation,

\begin{equation}
\hat{J}_z(+0) + \hat{J}_y(+0) = \hat{J}_z(-0) + \hat{J}_y(-0),
\label{eq-CurrentConservationGen}
\end{equation}
where $\hat{J}_z(z) = D \hat{g} \cdot \partial_z\hat{g}\big|_{y=0}$ and
$\hat{J}_y(y) = D \hat{g} \cdot \partial_y\hat{g} \big|_{z=0}$. We arrive at the equation

\begin{equation}
\kappa_{\omega}^2 L^2 \big[ \hat{\tau}_3, \hat{g} \big]
+
\left[ \left( r_z \hat{G}_z + r_y \hat{G}_y \right), \hat{g} \right]
= 0
\label{eq-UsadelIntegratedWithBoundaryConds}
\end{equation}
with $r_{\nu}=\kappa_{\nu}L$, $\hat{G}_{\nu} = \left[ \hat{g}_S \left(L_{\nu}\right) + \hat{g}_S\left(-L_{\nu}\right) \right]/2$.
Equation (\ref{eq-UsadelIntegratedWithBoundaryConds}) can be written in the form

\begin{equation}
\left[ \hat{M},\hat{g} \right]=0,
\label{eq-CommutatorM}
\end{equation}
where $\hat{M} = \kappa_{\omega}^2 L^2 \hat{\tau}_3 + \left( r_z \hat{G}_z +
r_y \hat{G}_y \right)$. The Green's functions in the S reservoirs have the form

\begin{equation}
\hat{g}_S \left( \zeta L_{\nu} \right) = g_S \hat{\tau}_3 + f_S \left[ \hat{\tau}_2
\cos(\varphi_{\nu \zeta})
+ \hat{\tau}_1 \sin(\varphi_{\nu \zeta}) \right],
\label{eq-gS}
\end{equation}
where $\varphi_{\nu\zeta}=\varphi_{\nu}(\zeta L_{\nu})$ is the phase of the order parameter in the superconductor at $z=\zeta L_z$ and $y=\zeta L_y$, respectively. For simplicity, we assume that the superconductors are made of
identical materials, i.e., their superconducting gap parameters $\Delta$ have equal magnitudes and therefore the functions

\begin{equation}
g_S = \frac{\omega}{\xi_{\omega}},
\quad
f_S = \frac{\Delta}{\xi_{\omega}},
\quad
\xi_{\omega} = \sqrt{\omega^{2}+\Delta^{2}}
\label{eq-g3}
\end{equation}
are identical for all superconductors. Using Eqs.~(\ref{eq-gS}) and (\ref{eq-g3}), the matrix $\hat{M}$ can be represented in the form

\begin{equation}
\hat{M} = \hat{\tau}_3 M_{3N} + f_S \left( \hat{\tau}_2 m_2 + \hat{\tau}_1 m_1 \right),
\label{eq-M-explicit}
\end{equation}
where $m_1 = r_z \sin\left(\Phi_{z}\right) \cos\left(\phi_{z}/2\right) + r_y
\sin\left(\Phi_{y}\right) \cos\left(\phi_{y}/2\right)$, $m_2 = r_z \cos\left(\Phi_{z}\right) \cos\left(\phi_{z}/2\right) + r_y  \cos\left(\Phi_{y}\right) \cos\left(\phi_{y}/2\right)$, $M_{3N} = [ \kappa_{\omega}^2 L^2 + \left( r_z + r_y \right) g_S ]$, $\Phi_{\nu} = (\varphi_{\nu+} + \varphi_{\nu-})/2$, and $\phi_{\nu} = \varphi_{\nu+} -
\varphi_{\nu-}$. The solution of this equation is sought in the form $\hat{g} = a\hat{M}$, where the coefficient $a$ can be determined from the normalization condition (\ref{eq-Norm}). We obtain $a^{-2} = \left[
M_{3N}^2 + f_S^2 (m_1^2 + m_2^2) \right]$, where the sign of $a$ is chosen in such a way that the density of states is a positive quantity. Thus, in the limit of a short Josephson junction, the solution of the Usadel equation is  given by

\begin{equation}
\hat{g}
=
\frac{\hat{M}}{\sqrt{M_{3N}^2 + f_S^2 (m_1^2 + m_2^2)}}
=
\frac{\hat{M}}{\mathcal{D}_N}
\label{eq-ResultForG}
\end{equation}
with $\mathcal{D}_N = \sqrt{M_{3N}^2 + f_S^2 (m_1^2 + m_2^2)}$.

Having determined the Green's function $\hat{g}$, we can now calculate physical observable quantities such as the density of states $N(\varepsilon)$ and the Josephson current $I_J$.

\section{Density of states}
\label{sec-dos}
\begin{figure}[!t]
\includegraphics[scale=0.25]{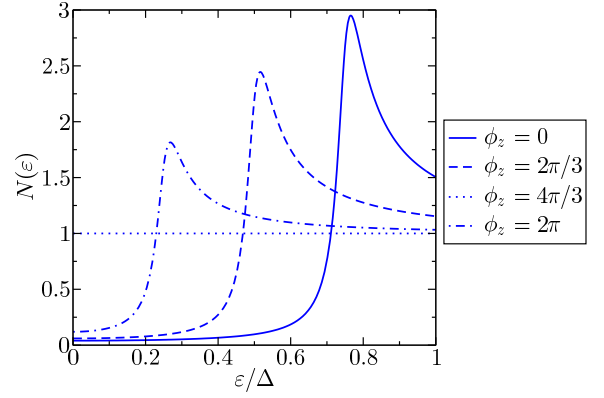}
\caption{(Color online) Density of states $N(\varepsilon)$ in the n$-$wire as a function of
normalized energy $\varepsilon/\Delta$ for $\phi_y = 2\pi/3$ and different phase differences
$\phi_z$. The plots are presented for the parameters $\gamma/\Delta = 0.03$, $r_z = r_y = 1$, and
$L/\xi_S = 0.1$.}
\label{fig-DOS}
\end{figure}
With help of the solution for the Green's function $\hat{g}$, Eq.~(\ref{eq-ResultForG}), we can find the density of states $N(\varepsilon)$ in the n$-$wire,

\begin{equation}
N(\varepsilon)
=
\mathrm{Re}\left[ \left. \frac{M_{3N}}{\mathcal{D}_N} \right|_{\omega \rightarrow -i(\varepsilon +
i0)} \right].
\label{eq-DOS}
\end{equation}
One can consider the following limiting cases of Eq.~(\ref{eq-DOS}).

For very high interface resistances $R_{nS,\nu} \gg R_L$ (i.e., $r_{\nu} \ll 1$), where $R_L = L\rho/2$, one can neglect the term containing $f_S$ in the definition of $\mathcal{D}_N$ so that $N(\varepsilon) \approx 1$. For very low interface resistances $R_{nS,\nu} \ll R_L$ (i.e., $r_{\nu} \gg 1$) we set for simplicity $\kappa_z = \kappa_y = \kappa$, i.e., $r_z = r_y = r$. Then, $M_{3N} \approx 2r g_S$ and $\mathcal{D}_N \approx 2r [g_S^2 + \left(f_S/2\right)^2 \left( \mu_1^2 + \mu_2^2 \right)]^{1/2}$ with
$\mu_1 = \sin(\Phi_{z}) \cos(\phi_{z}/2) + \sin(\Phi_{y}) \cos(\phi_{y}/2)$ and $\mu_2 = \cos(\Phi_{z}) \cos(\phi_{z}/2) + \cos(\Phi_{y}) \cos(\phi_{y}/2)$. This results in $N(\varepsilon) \approx \mathrm{Re}\left[ \varepsilon/\sqrt{\varepsilon^2 - \Delta^2 \left( \mu_1^2 + \mu_2^2 \right)/4} \right]$, i.e., we obtain a gap induced in the n$-$wire equal to $\Delta $ in the absence of superconducting phases ($\Phi_{\nu} = \phi_{\nu} = 0$) and turning to zero for $\mu_1 = \mu_2 = 0$.

In the limit of small energies ($\varepsilon/\Delta \ll 1$) and small values
of $r$, we obtain $N(\varepsilon) \approx \mathrm{Re}\left[  \varepsilon/\sqrt{\varepsilon^2 - E_{\mathrm{th}}^2 r^2 \left( \mu_1^2 + \mu_2^2 \right)} \right]$. At $\Phi_{\nu} = \phi_{\nu} = 0$, i.e., at $\mu_1 = 0$ and $\mu_2 = 2$, this expression coincides with McMillan's formula
\cite{McMillan}.

In Fig.~\ref{fig-DOS}, we show the dependence of the density of states $N(\varepsilon)$ on the normalized energy $\varepsilon/\Delta$ for different phase differences $\phi_z$ keeping $\phi_y$ fixed at $2\pi/3$. Here, we have set $\varphi_{\nu+} = -\varphi_{\nu-} = \phi_{\nu}/2$ so that $\Phi_{\nu} = 0$ and $m_1 = 0$. Moreover, since the DOS is calculated from  $N(\varepsilon)=\mathrm{Re}\left[\hat{\tau}_3 \cdot \hat{g}^R
\right]$, i.e., from the retarded Green's function, we replace in Eq.~(\ref{eq-DOS}) $\varepsilon/\Delta \rightarrow (\varepsilon + i\gamma)/\Delta$ and choose a damping $\gamma/\Delta=0.03$
according to analyticity of $\hat{g}^R$ in the upper half plane.

As already mentioned above, we find a gap induced in the n$-$wire. The position of the gap with respect to normalized energy $\varepsilon/\Delta$ can be tuned by variation of the parameters $r_{\nu}$ and the superconducting phase differences $\phi_{\nu}$. In particular, it can be seen that
for fixed phase difference $\phi_{\nu}$ one can always find a phase difference $\phi_{\bar{\nu}} = 2\arccos[-r_{\bar{\nu}} \cos(\phi_{\bar{\nu}}/2)/r_{\nu}]$ such that the density of states becomes unity for all energies, i.e., the gap turns to zero. Here, $\bar{\nu}=y,z$ denotes the axis perpendicular to $\nu=z,y$, i.e., if, for instance, $\nu=z$, then $\bar{\nu}=y$. Since $N(\varepsilon)$ is only dependent on $\cos(\phi_z/2)$ due to symmetry, it is sufficient to consider phase differences $\phi_z \in [0, 2\pi]$ for constant $\phi_y$.

The dependence of the DOS on the phase difference between the superconducting leads in S/N/S systems has been studied in many papers (see, e.g., Ref.~\onlinecite{AslamVolkov86} as well as Ref.~\onlinecite{Bezug03} and references therein). Indeed, the proximity induced gap in the n$-$wire has also been obtained in Ref.~\onlinecite{Bezug03} for a three-terminal
diffusive S/N/S junction with normal metal wires. In agreement with our results, the authors of Ref.~\onlinecite{Bezug03} report a proximity induced gap in the DOS of the n$-$wire which depends on the superconducting phase difference $\phi$.

\begin{figure}[t]
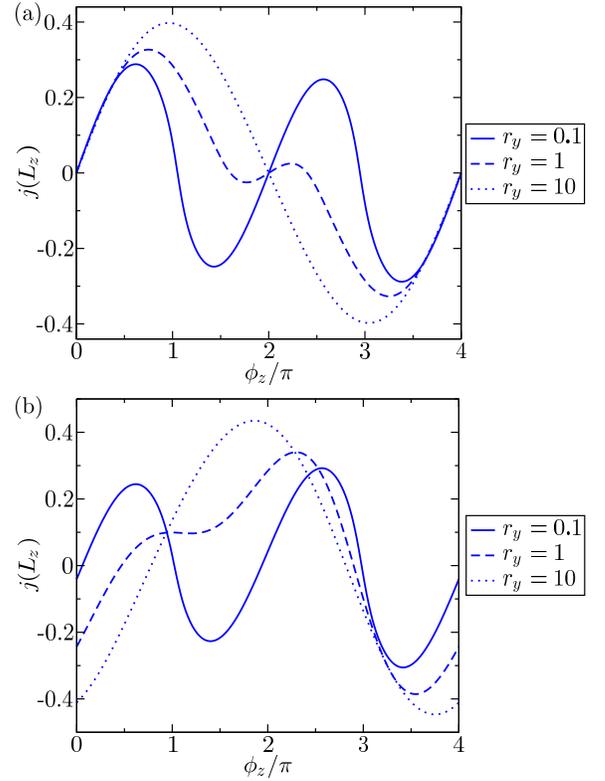

\includegraphics[scale=0.25]{Fig3a-Jz_N1_rz1}\vspace{1mm}
\includegraphics[scale=0.25]{Fig3b-Jz_N2_rz1}
\caption{(Color online) Normalized supercurrent $j(L_z)$ in the horizontal wire at the right interface as a function of the horizontal phase difference $\phi_z$ for different values of the resistance ratio $r_y=R_L/R_{nS,y}$. The plots are presented for the two cases (a) $\varphi_{\nu +} = -\varphi_{\nu -} = \phi_{\nu}/2 \, [\Phi_{\nu}=0]$ with $\nu=z,y$ and (b) $\varphi_{z+}=-\varphi_{z-}=\phi_z/2 \, [\Phi_z=0]$,  $\varphi_{y+}=\varphi_{y-}=\Phi_y \, [\phi_y=0]$. The parameters are $r_z = 1$, $T/\Delta=0.05$, $L/\xi_S=0.1$, (a) $\phi_y=2\pi/5$, and (b) $\Phi_y=2\pi/5$.}
\label{fig-JosephsonCurrent-rY}
\end{figure}

\section{Josephson current}
\label{sec-Jc}
By means of Eq.~(\ref{eq-ResultForG}), we can also calculate the Josephson current flowing through the four terminals located at $\nu = \zeta L_{\nu}$ given by

\begin{equation}
I_J(\zeta L_{\nu})
=
\zeta \frac{i\pi T}{4eR_{nS,\nu}} \sum_{\omega} \mathrm{Tr}\left\{ \hat{\tau}_3 \cdot \left[
\hat{g}, \hat{g}_S \right] \right\}
\Big|_{\nu=\zeta L_{\nu}},
\label{eq-supercurrent-full}
\end{equation}
where Tr is the trace over particle-hole indices. Substituting the expressions for $\hat{g}$ and $\hat{g}_S$, Eqs.~(\ref{eq-ResultForG}) and (\ref{eq-gS}), into this expression, we obtain for the normalized supercurrents $j(\zeta L_{\nu}) \equiv I_J(\zeta L_{\nu})/I_{J0\nu}$ [where $I_{J0\nu} = \pi\Delta/(eR_{nS,\nu})$]

\begin{figure*}[!t]
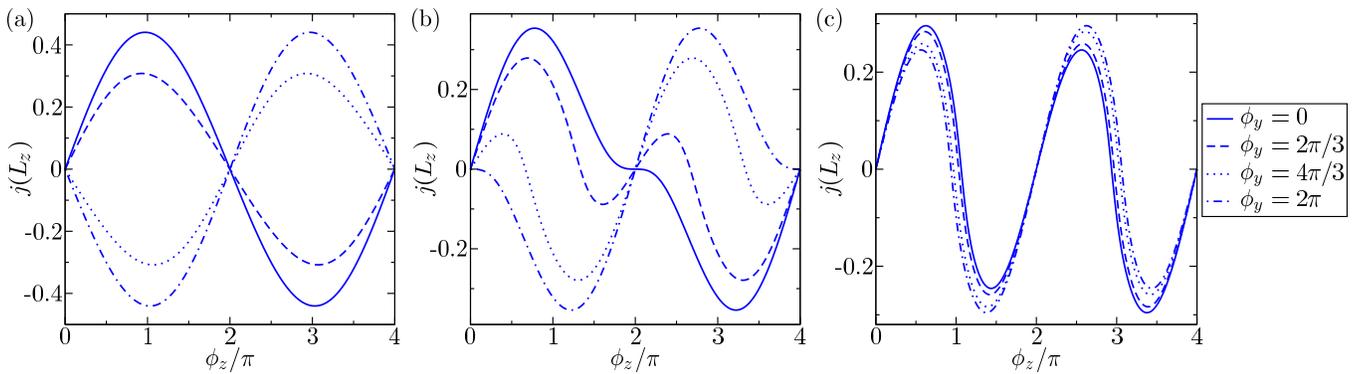

\includegraphics[scale=0.238]{Fig4a-Jz_N1_rz1-ry10}
\includegraphics[scale=0.238]{Fig4b-Jz_N1_rz1-ry1}
\includegraphics[scale=0.238]{Fig4c-Jz_N1_rz10-ry1}
\caption{(Color online) Normalized supercurrent $j(L_z)$ in the horizontal wire at the right interface as a function of the horizontal phase difference $\phi_z$ for different values of the vertical phase difference $\phi_y$. The plots are presented for the case $\varphi_{\nu +} = -\varphi_{\nu -} = \phi_{\nu}/2 \, [\Phi_{\nu}=0]$ with $\nu=z,y$ and the parameters $T/\Delta=0.05$, $L/\xi_S=0.1$, (a) $r_z = 1$, $r_y = 10$, (b) $r_z = r_y = 1$, and (c) $r_z = 10$, $r_y = 1$.}
\label{fig-JosephsonCurrent-PhiY}
\end{figure*}
\begin{eqnarray}
j(\zeta L_{\nu})
&=&
\frac{T}{\Delta}
\Big[
\frac{r_{\nu}}{2} \sin\left( \phi_{\nu} \right) + \zeta r_{\bar{\nu}} \cos\left( \phi_{\bar{\nu}}/2
\right)
\nonumber
\\
&&
\phantom{\frac{T}{\Delta}\Big[}
\times
\sin\left( \varphi_{\nu\zeta} - \Phi_{\bar{\nu}} \right)
\Big]
\sum_{\omega}
\frac{\Delta^2}{\omega^2 + \Delta^2} \frac{1}{\mathcal{D}_N}.
\quad
\label{eq-supercurrent}
\end{eqnarray}
This formula determines the Josephson current between identical  superconductors (a generalization to the case of different superconductors is straightforward) for arbitrary phases of the order parameter and arbitrary strength of proximity effect, i.e., the parameters $r_{\nu}$ are not restricted to particular limits. It is obvious that, as it should be, a shift of all the phases by a constant does not change the currents $j(\zeta  L_{\nu})$. In the following we consider two particular cases with respect to the superconducting phases $\varphi_{\nu\zeta}$.

First, we assume that $\varphi_{\nu+} = -\varphi_{\nu-} = \phi_{\nu}/2$ so that $\Phi_{\nu} = 0$. Then, one can easily obtain from  Eq.~(\ref{eq-supercurrent}) the Josephson current in the horizontal and vertical branches,

\begin{eqnarray}
j(\zeta L_{\nu})
&=&
\frac{T}{\Delta}
\Big[
\frac{r_{\nu}}{2} \sin\left( \phi_{\nu} \right) + r_{\bar{\nu}} \cos\left( \phi_{\bar{\nu}}/2
\right)
\nonumber
\\
&&
\phantom{\frac{T}{\Delta}\Big[}
\times
\sin\left( \phi_{\nu}/2 \right)
\Big]
\sum_{\omega}
\frac{\Delta^2}{\omega^2 + \Delta^2} \frac{1}{\mathcal{D}_N}.
\label{eq-supercurrent-case1}
\end{eqnarray}
One can see that if $R_{nS,\nu} \ll R_{nS,\bar{\nu}}$, i.e., $r_{\nu} \gg r_{\bar{\nu}}$, the dependence of the Josephson current $j(L_{\nu})$ on the phase difference $\phi_{\nu}$ is close to the sinusoidal one with the period $2\pi$ (note that in the considered particular case $m_1=0$, but the function $m_2 \neq 0$ and it also depends on $\phi_{\nu}$). In the opposite limit $R_{nS,{\nu}} \gg R_{nS,\bar{\nu}}$, i.e., $r_{\nu} \ll r_{\bar{\nu}}$, the phase dependence of the current $j(L_{\nu})$ is changed drastically as it is almost sinusoidal but the period of oscillations is equal to $4\pi$.

The obtained phase dependence of the Josephson current leads to an  interesting phenomenon. As is well known (see, for example, Refs.~\onlinecite{Kulik,Likharev,Barone}), if a Josephson junction is irradiated by microwaves with frequency $\omega$ and if a bias voltage $V$ is applied between the superconducting leads the so-called Shapiro steps appear on the current-voltage characteristics $I(V)$. The positions of the Shapiro steps are determined by the relation $V_{\mathrm{Sh},n} = \hbar\omega/2en$ with $n=1,2,\ldots$. Thus, the main Shapiro step corresponds to the voltage $V_{\mathrm{Sh},1} = \hbar\omega/2e$. However, given a bias voltage $V$ is applied between the superconducting leads in the horizontal wire and
the phases are varying in time with frequencies much less than the characteristic energies ($\Delta$ or $T_c$) of the problem, one can
conclude from Eq.~(\ref{eq-supercurrent-case1}) that for $r_z \ll r_y$ the positions of the Shapiro steps are changed: $V_{\mathrm{Sh},n}(r_z \ll r_y) = \hbar\omega/en$. Moreover, the positions of the Shapiro steps depend on the phase difference $\phi_y$, that is, on the current flowing through the  vertical wire. Note that the $4\pi$ periodic phase dependence of the Josephson current is typical for junctions with Majorana bound states \cite{Majorana1,Majorana2,Majorana3,Majorana4}.

Second, consider now the case when in the horizontal branch $\varphi_{z+} = -\varphi_{z-} = \phi_z/2$, i.e., there is no change compared to the previous case, but in the vertical branch $\varphi_{y+} = \varphi_{y-} = \Phi_y$ $(\phi_y = 0)$. Then, we find

\begin{eqnarray}
j(\zeta L_z)
&=&
\frac{T}{\Delta}
\left[
\frac{r_z}{2} \sin\left( \phi_z \right) + r_y \sin\left( \phi_z/2 - \zeta \Phi_y \right)
\right]
\nonumber
\\
&&
\times
\sum_{\omega}
\frac{\Delta^2}{\omega^2 + \Delta^2} \frac{1}{\mathcal{D}_N}
\label{eq-supercurrent-case2-z}
\end{eqnarray}
and

\begin{eqnarray}
j(\zeta L_y)
&=&
\zeta \frac{T}{\Delta}
r_z \cos\left( \phi_z/2 \right) \sin\left( \Phi_y \right)
\nonumber
\\
&&
\times
\sum_{\omega}
\frac{\Delta^2}{\omega^2 + \Delta^2} \frac{1}{\mathcal{D}_N}.
\label{eq-supercurrent-case2-y}
\end{eqnarray}
In this case the currents through the interfaces at $z = \zeta L_z$ are different if $r_y \gg r_z$. The currents through the interfaces at $y = \zeta L_y$ flow in opposite directions.

In Fig.~\ref{fig-JosephsonCurrent-rY}, we plot the dependence of the current $j(L_z)$ on the horizontal phase difference $\phi_z$ for different values of the vertical resistance ratio $r_y$ and fixed horizontal resistance ratio $r_z = 1$. The curves are presented for the cases $\varphi_{\nu+}=-\varphi_{\nu-}=\phi_{\nu}/2$ with $\phi_y = 2\pi/5$,
Fig.~\ref{fig-JosephsonCurrent-rY}(a), and  $\varphi_{z+}=-\varphi_{z-}=\phi_z/2$, $\varphi_{y+} = \varphi_{y-} = \Phi_{y} = 2\pi/5$, Fig.~\ref{fig-JosephsonCurrent-rY}(b).

From Fig.~\ref{fig-JosephsonCurrent-rY}(a), it can be nicely seen how the phase dependence of the supercurrent $j(L_z)$ changes from a $4\pi$  periodicity for $r_y \gg r_z$ (dotted curve) to a $2\pi$ periodicity for $r_y \ll r_z$ (solid curve). In the intermediate region where the parameters $r_z$ and $r_y$ are of the same order of magnitude (dashed curve), we find two local extremes close to $\phi_z = 2\pi$.

From Fig.~\ref{fig-JosephsonCurrent-rY}(b), we conclude that for $r_z \gg r_y$ (solid curve) there also emerges a $2\pi$ periodicity that, compared to  Fig.~\ref{fig-JosephsonCurrent-rY}(a), is slightly shifted in the horizontal direction due to the different dependencies of the second term in Eqs.~(\ref{eq-supercurrent-case1}) and (\ref{eq-supercurrent-case2-z}) on the phase difference $\phi_y$ and the phase sum $\Phi_y$, respectively. Decreasing the parameter $r_y$ further results in a perfect sinusoidal dependence on $\phi_z$. Here, too, in the limit $r_z \ll r_y$ (dotted curve), the periodicity of the supercurrent $j(L_z)$ is $4\pi$ but as for the case $r_z \gg r_y$, it is shifted in the horizontal direction. For the intermediate region $r_z \approx r_y$ (dashed curve), we find a different influence of the resistance ratios $r_{\nu}$ than in Fig.~\ref{fig-JosephsonCurrent-rY}(a).
In this case, we find a double peak rather than two local extremes.

\begin{figure}[!t]
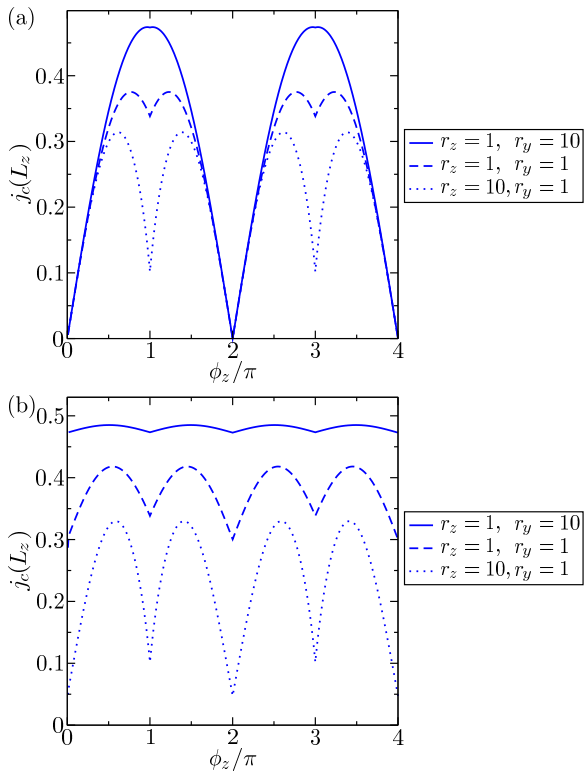

\includegraphics[scale=0.25]{Fig5a-JzCr_N1}\vspace{1mm}
\includegraphics[scale=0.25]{Fig5b-JzCr_N2}
\caption{(Color online) Normalized critical supercurrent $j_c(L_z)$ in the horizontal wire at the right interface as a function of the phase difference $\phi_z$ for different values of the resistance ratios $r_{\nu}$ where $\nu=z,y$. The plots are presented for the two cases (a) $\varphi_{\nu +} = -\varphi_{\nu -} = \phi_{\nu}/2 \, [\Phi_{\nu}=0]$ and (b) $\varphi_{z+}=-\varphi_{z-}=\phi_z/2 \, [\Phi_z=0]$,  $\varphi_{y+}=\varphi_{y-}=\Phi_y \, [\phi_y=0]$. The parameters are $T/\Delta=0.05$ and $L/\xi_S=0.1$.}
\label{fig-CriticalJosephsonCurrent}
\end{figure}
In Fig.~\ref{fig-JosephsonCurrent-PhiY}, we show the dependence of the supercurrent $j(L_z)$ in the horizontal wire at the right interface on the horizontal phase difference $\phi_z$ for different values of the vertical phase difference $\phi_y$ and the resistance ratios $r_{\nu}$. Here, we only
analyze the case $\varphi_{\nu +} = - \varphi_{\nu -} = \phi_{\nu}/2$.

As can be seen from Fig.~\ref{fig-JosephsonCurrent-PhiY}(a), where the case $r_z \ll r_y$ is displayed, the supercurrent $j(L_z)$ has a perfect sinusoidal shape with a $4\pi$ periodicity for arbitrary vertical phase
differences $\phi_y$. This can be understood by means of  Eq.~(\ref{eq-supercurrent-case1}) where for $r_z \ll r_y$ the first term in the square brackets can be neglected compared to the second term so that the current is proportional to $\cos(\phi_y/2) \sin(\phi_z/2)$. Therefore, due to the dependence on $\cos(\phi_y/2)$, the variation of the vertical phase difference $\phi_y$ from 0 to $2\pi$ leads to a sign change of $j(L_z)$.

In the opposite limit $r_z \gg r_y$, which is shown in  Fig.~\ref{fig-JosephsonCurrent-PhiY}(c), the first term in the square brackets of Eq.~(\ref{eq-supercurrent-case1}) dominates with respect to the
second term, i.e., in this case, the supercurrent $j(L_z)$ is proportional to $\sin(\phi_z)$ and is independent of the vertical phase difference $\phi_y$. Thus, the curves collapse into a single curve with periodicity $2\pi$.

In the intermediate regime $r_z \approx r_y$ the form of supercurrent $j(L_z)$ can be considerably changed by variation of the vertical phase difference $\phi_y$. While the dependence of the current on the horizontal phase difference $\phi_z$ for $\phi_y=0, 2\pi$ is nearly sinusoidal, we find a clearly distinct behavior for $\phi_y = 2\pi/3, 4\pi/3$ where the dependence on $\phi_z$ is more complicated.

Finally, we calculate numerically the normalized critical Josephson current $j_c(L_z)$ in the horizontal wire as a function of the horizontal phase difference $\phi_z$ for different values of the resistance ratios $r_{\nu}$. For each set of parameters $r_{\nu}$, we calculate for each phase difference $\phi_z \in [0,4\pi]$ the maximal supercurrent $j_c(L_z)$ as a function of $\phi_y$ and plot it versus $\phi_z$ as shown in  Fig.~\ref{fig-CriticalJosephsonCurrent}. Interestingly, we obtain for the case $\varphi_{\nu +} = -\varphi_{\nu -} = \phi_{\nu}/2$, Fig.~\ref{fig-CriticalJosephsonCurrent}(a), two maximal values of $j_c(L_z)$ for the case $r_z \ll r_y$ and four maxima of $j_c(L_z)$ for the case
$r_z \gg r_y$. In contrast to Fig.~\ref{fig-CriticalJosephsonCurrent}(a), in the case where we have in the vertical wire  $\varphi_{y+}=\varphi_{y-} =\Phi_y$, Fig.~\ref{fig-CriticalJosephsonCurrent}(b), we always find four maximal values of $j_c(L_z)$ and there does not exist a horizontal phase
difference $\phi_z$ for which the critical current vanishes.

In Fig.~\ref{fig-CriticalJosephsonCurrent}, the number of peaks of the critical current reflects the periodicity of the Josephson current with respect to $\phi_z$ which can be understood by considering Eqs.~(\ref{eq-supercurrent-case1}) and (\ref{eq-supercurrent-case2-z}), respectively. Considering the upper panel of  Fig.~\ref{fig-CriticalJosephsonCurrent}, we find for the case $r_z \ll r_y$ (solid curve) two peaks of the critical current because according to Eq.~(\ref{eq-supercurrent-case1}) the prefactor of the Josephson current $j(L_z)$ in this limit is proportional to $\sin(\phi_z/2)$, i.e., it has a $4\pi$ periodicity [also see Fig.~\ref{fig-JosephsonCurrent-PhiY}(a)]. On the  contrary, for $r_z \gg r_y$ (dotted curve), we obtain $j(L_z) \propto \sin(\phi_z)$, i.e., the Josephson current has a $2\pi$ periodicity resulting in four peaks of the according critical current. Similarly, we can explain the appearance of four peaks in Fig.~\ref{fig-CriticalJosephsonCurrent}(b). Here, in the limit $r_z \ll r_y$ (solid curve), the Josephson current is proportional to $\sin(\phi_z/2-\Phi_y)$ and in the opposite limit (dotted curve) it is proportional to $\sin(\phi_z)$. Therefore, in the latter case, we expect four peaks of $j_c(L_z)$, whereas for $r_z \ll r_y$ the critical current approaches a constant as a function of $\phi_z$ because, here, the superconducting phase difference $\phi_z$ only leads to a horizontal shift of the Josephson current and does not affect its magnitude. According to Eq.~(\ref{eq-ShortJunctionCond}), where the condition $L_{\nu}^2 \ll D/\Delta=\xi_S^2$ is supposed to be fulfilled,  Figs.~\ref{fig-JosephsonCurrent-rY}--\ref{fig-CriticalJosephsonCurrent}
are presented for a small value of normalized junction length $L/\xi_S$, which is kept fixed at $L/\xi_S=0.1$.

\section{Ferromagnetic Wires with Non-collinear Magnetization}
\label{sec-Fwire}
In this section, we consider the case when the cross is formed not by
normal wires but by ferromagnetic wires with non-collinear magnetization
vectors $\mathbf{M}$. We assume that the vector $\mathbf{M}$ is aligned in the $z$ direction in the horizontal ferromagnetic wire and along the $y$ direction in the vertical ferromagnetic wire, i.e., in both cases $\mathbf{M}$ points along the corresponding wire. In order to describe the condensate in this case, the quasiclassical Green's functions should be generalized. They become $4\times 4$ matrices $\check{g}$ in the product space of particle-hole and spin variables (see, e.g.,  Ref.~\onlinecite{BVErmp}). Equation~(\ref{eq-UsadelEqn}) can be written as

\begin{align}
-\partial_z ( \check{g} \cdot \partial_z \check{g} ) + \kappa_{\omega}^2
[ \hat{\tau}_3 \cdot \hat{\sigma}_0, \check{g} ] - i\kappa_F^2 [\hat{\tau}_3 \cdot \hat{\sigma}_3,
\check{g} ]
&=
0,
\label{eq-Usadel-SF-z}
\\
-\partial_y ( \check{g} \cdot \partial_y \check{g} ) + \kappa_{\omega}^2
[ \hat{\tau}_3 \cdot \hat{\sigma}_0, \check{g} ] - i\kappa_F^2 [\hat{\tau}_0 \cdot
\hat{\sigma}_2,\check{g} ]
&=
0,
\label{eq-Usadel-SF-y}
\end{align}
where $\kappa_F^2 = V_{\mathrm{ex}}/D$ and $V_{\mathrm{ex}}$ is the exchange energy. The matrices $\hat{\tau}_i$ and $\hat{\sigma}_i$ operate in the particle-hole and spin spaces, respectively (Pauli matrices for $i = 1,2,3$ and identity matrices for $i=0$). We assume that the boundary condition (\ref{eq-BoundaryConds}) is valid in the generalized form ($\hat{g}\to \check{g}$, $\hat{g}_S\to \check{g}_S$) also in the considered case, i.e., we
assume that the S/F interfaces do not affect the spins.

Following the procedure presented in Sec.~\ref{sec-model}, we obtain a generalized version of Eq.~(\ref{eq-CommutatorM}),

\begin{equation}
\left[ \check{M}_F, \check{g} \right] = 0
\label{eq-CommutatorMF}
\end{equation}
with $\check{M}_F$ given by

\begin{eqnarray}
\check{M}_F(\alpha)
&=&
\kappa_{\omega}^2 L^2 \, \hat{\tau}_3 \cdot \hat{\sigma}_0
-
i v_{\mathrm{ex}} \left[
\sin(\alpha) \hat{\tau}_0 \cdot \hat{\sigma}_2 + \cos(\alpha) \hat{\tau}_3 \cdot \hat{\sigma}_3
\right]
\nonumber
\\
&&
+
r_y \check{G}_y + r_z \check{G}_z,
\end{eqnarray}
where $v_{\mathrm{ex}}=\kappa_{F}^2 L \sqrt{L_z^2+L_y^2}$ is the normalized exchange energy, $\tan(\alpha) = L_y/L_z$, and $\check{G}_{\nu} = \left[\check{g}_S \left(L_{\nu}\right) + \check{g}_S \left(-L_{\nu}\right) \right]/2$ with

\begin{align}
\check{g}_{S}(\zeta L_{\nu})
&=
g_S \hat{\tau}_3 \cdot \hat{\sigma}_0
\nonumber
\\
&\quad
+ f_S
\left[ \hat{\tau}_2 \cos(\varphi_{\nu\zeta}) + \hat{\tau}_1 \sin(\varphi_{\nu\zeta}) \right]
\cdot \hat{\sigma}_3.
\label{eq-gScheck}
\end{align}
As before, Eqs.~(\ref{eq-Usadel-SF-z}) and (\ref{eq-Usadel-SF-y}) must be solved together with the normalization condition

\begin{equation}
\check{g} \cdot \check{g} = \check{\mathds{1}}.
\label{eq-NormalizationCondCheck}
\end{equation}
At $V_{\mathrm{ex}} = 0$, the matrix element $(\check{g})_{11}$ in the spin space coincides with the matrix $\hat{g}$ given by  Eq.~(\ref{eq-ResultForG}).

One can exclude the term proportional to $\hat{\tau}_0 \cdot \hat{\sigma}_2$ in the matrix $\check{M}_{F}$ by making a transformation (see Ref.~\onlinecite{BVErmp}) that describes a rotation in the spin space,

\begin{equation}
\check{\mathcal{M}}_{F}
=
\check{U}^{\dagger} \cdot \check{M}_{F} (\alpha) \cdot \check{U},
\quad
\check{\mathfrak{g}} = \check{U}^{\dagger} \cdot \check{g} \cdot \check{U},
\label{eq-TransformationByU}
\end{equation}
where the transformation matrix depends on $\alpha$ via $\check{U}(\alpha) = \exp(i\hat{\tau}_3 \cdot \hat{\sigma}_1 \alpha/2) = \cos(\alpha/2) + i\sin(\alpha/2) \hat{\tau}_3 \cdot \hat{\sigma}_1$. One can see that the matrices $\check{G}_{\nu}$ remain unchanged (they commute with $\check{U}$)
because they describe the singlet superconductors that serve as reservoirs and are not affected by a rotation in the spin space. Similarly, the first term in $\check{M}_{F}(\alpha)$ commutes with $\check{U}$. Therefore, after this transformation, Eq.~(\ref{eq-CommutatorMF}) attains the form

\begin{equation}
[\check{\mathcal{M}}_{F}, \check{\mathfrak{g}}] = 0,
\label{eq-CommutatorMFtransformed}
\end{equation}
where $\check{\mathcal{M}}_{F}=\check{M}_F(0)$ is given by

\begin{equation}
\check{\mathcal{M}}_{F}
=
\kappa_{\omega}^2 L^2\, \hat{\tau}_3 \cdot \hat{\sigma}_0
-
i v_{\mathrm{ex}} \hat{\tau}_3 \cdot \hat{\sigma}_3
+
r_z \check{G}_z + r_y \check{G}_y.
\end{equation}
Due to the diagonal structure of $\check{\mathcal{M}}_{F}$ in the spin space
the equations for different spin sectors decouple, and the diagonal elements $\hat{\mathfrak{g}}_{s}$ ($s = \pm$, $\hat{\mathfrak{g}}_{+}=\hat{\mathfrak{g}}_{11}$, $\hat{\mathfrak{g}}_{-}=\hat{\mathfrak{g}}_{22}$) of $\check{\mathfrak{g}}$ can be obtained as it was done in Sec.~\ref{sec-model}. This yields

\begin{equation}
\hat{\mathfrak{g}}_{s} = \frac{\hat{\mathcal{M}}_{Fs}}{\mathcal{D}_{Fs}}
\end{equation}
with

\begin{align}
\hat{\mathcal{M}}_{Fs}
&=
\hat{\tau}_3 \left( M_{3N} - s iv_{\mathrm{ex}} \right) + s f_S \left( \hat{\tau}_2 m_2 +
\hat{\tau}_1 m_1 \right),\\
\mathcal{D}_{Fs}
&=
\sqrt{\left(M_{3N} - s iv_{\mathrm{ex}} \right)^2 + f_S^2 \left(m_1^2 + m_2^2 \right)}
\label{eq-Def-MFns}
\end{align}
satisfying $\mathcal{D}_{F-} = \mathcal{D}_{F+}^{\ast}$ for real $\omega$. The resulting matrix $\check{\mathfrak{g}}$ can be written as

\begin{equation}
\check{\mathfrak{g}} = \hat{\mathfrak{g}}_{0} \cdot \hat{\sigma}_0 + \hat{\mathfrak{g}}_{3} \cdot
\hat{\sigma}_3,
\label{eq-CheckgFromCheckgTranformed}
\end{equation}
where $\hat{\mathfrak{g}}_{0,3} = \left( \hat{\mathfrak{g}}_{+} \pm \hat{\mathfrak{g}}_{-} \right)/2$. Exploiting the representation of $\check{\mathcal{M}}_{F}$ in terms of its diagonal blocks $\hat{\mathcal{M}}_{Fs}$ as

\begin{equation}
\check{\mathcal{M}}_{F} = \sum_s \hat{\mathcal{M}}_{Fs} \cdot \tfrac12 (\hat{\sigma}_{0} +
s\hat{\sigma}_{3}),
\end{equation}
we arrive at

\begin{equation}
\check{\mathfrak{g}} = \check{\mathcal{M}}_{F}
\left(
\operatorname{Re}\left[\mathcal{D}_{F+}^{-1}\right]\hat{\sigma}_0 + i
\operatorname{Im}\left[\mathcal{D}_{F+}^{-1}\right]\hat{\sigma}_3
\right).
\end{equation}
Using Eq.~(\ref{eq-TransformationByU}) one can obtain the matrix $\check{g} = \check{U} \cdot \check{\mathfrak{g}} \cdot \check{U}^{\dagger}$. Since $\check{\mathcal{M}}_{F}$ simply transforms back to $\check{M}_{F}$ it is sufficient to transform $\hat{\sigma}_3$ and we obtain

\begin{align}
\check{g} = \check{M}_{F}
\Big(
&\,\mathrm{Re}   \left[\mathcal{D}_{F+}^{-1}\right]\hat{\sigma}_0 \label{eq-gcheckfinal} \\
+ i &\,\mathrm{Im}   \left[\mathcal{D}_{F+}^{-1}\right] \left\{ \cos(\alpha)
\hat{\tau}_0\cdot\hat{\sigma}_3 + \sin(\alpha) \hat{\tau}_3\cdot\hat{\sigma}_2 \right\}
\Big).
\nonumber
\end{align}
Note that the real and imaginary parts of $\mathcal{D}_{F+}^{-1}$ present in the equations above must be evaluated treating $\omega$ as real before  applying analytical continuation via, e.g.,

\begin{equation}
N(\varepsilon) = \mathrm{Re}
\left[ \tfrac14 \operatorname{Tr}
\left\{
(\hat{\tau}_3 \cdot \hat{\sigma}_0) \cdot\check{g}
\right\}\Big\vert_{\omega\to -i(\varepsilon+i0)}
\right],
\end{equation}
where $\operatorname{Tr}$ is the trace over particle-hole and spin indices.

We are interested in the condensate function $\check{f}$ (the part of $\check{g}$ non-diagonal in the particle-hole space) that has the form
$\check{f} = \check{f}_{\mathrm{sg}} + \check{f}_{\mathrm{tr}}$ with the singlet component

\begin{equation}
\check{f}_{\mathrm{sg}}
=
\mathrm{Re}\left[ \mathcal{D}_{F+}^{-1}\right] f_S \left( \hat{\tau}_2 m_2 + \hat{\tau}_1 m_1
\right) \cdot \hat{\sigma}_3
\label{eq-fsng}
\end{equation}
and the triplet component

\begin{widetext}
\begin{equation}
\check{f}_{\mathrm{tr}}
=
i \, \mathrm{Im}\left[ \mathcal{D}_{F+}^{-1} \right] f_S \left\{
\left[ \hat{\tau}_2 \cdot \hat{\sigma}_0 \cos(\alpha) + \hat{\tau}_1 \cdot \hat{\sigma}_1
\sin(\alpha) \right] m_2
+
\left[ \hat{\tau}_1 \cdot \hat{\sigma}_0 \cos(\alpha) - \hat{\tau}_2 \cdot \hat{\sigma}_1
\sin(\alpha) \right] m_1
\right\}.
\label{eq-ftr}
\end{equation}
\end{widetext}
In Eq.~(\ref{eq-ftr}) the terms proportional to $\hat{\sigma}_{0}$ describe the triplet component with zero projection of the total spin of the Cooper pair on the $z-$axis ($\check{f}_{\mathrm{tr},0}$) whereas the terms proportional to $\hat{\sigma}_1$ describe the component with non-zero projection of the total spin of the Cooper pair on the $z-$axis ($\check{f}_{\mathrm{tr},1}$). The appearance of this ``$S_z\neq0$"-triplet component, which corresponds to a long-range triplet component in long ferromagnets (see Ref.~\onlinecite{BVErmp}), is easily understood from the physical point of view. In the absence of the horizontal wire, only singlet and triplet components with zero projection on the $y$ axis ($S_y = 0$) are induced in the vertical wire due to proximity effect. On the other hand, this triplet component has a nonzero projection on the $z$ axis ($S_z \neq 0$).

\begin{figure}[!t]
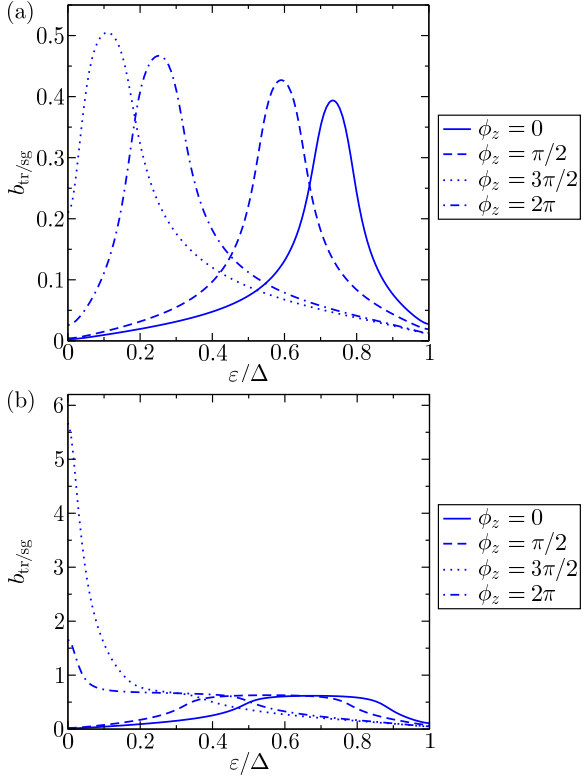

\includegraphics[scale=0.25]{Fig6a-TrSgRatio_F1_v5}
\includegraphics[scale=0.25]{Fig6b-TrSgRatio_F1_v20}
\caption{(Color online) Ratio of the ``$S_z\neq0$"-triplet component amplitude to that of the singlet component as a function of normalized energy  $\varepsilon/\Delta$ for the case $\varphi_{\nu+} = -\varphi_{\nu-} = \phi_{\nu}/2$ $(\Phi_{\nu}=0)$, $\phi_y=2\pi/3$, and different phase differences $\phi_z$. The plots display the function  $b_{\mathrm{tr}/\mathrm{sg}}$ for $\gamma/\Delta = 0.03$, $r_z=r_y=1$, $L_z/\xi_S=L_y/\xi_S=0.1$, and (a) $V_{\mathrm{ex}}/\Delta=5$, (b) $V_{\mathrm{ex}}/\Delta=20$.}
\label{fig-TrSgRatio}
\end{figure}

In Fig.~\ref{fig-TrSgRatio}, we plot the ratio of the ``$S_z\neq0$"-triplet component amplitude to that of the singlet component, i.e.,  $b_{\mathrm{tr}/\mathrm{sg}} = \vert \sin(\alpha) \mathrm{Im}[\mathcal{D}_{F+}^{-1}]/\mathrm{Re}[\mathcal{D}_{F+}^{-1}] \vert$, as a function of normalized energy $\varepsilon/\Delta$ for the case $\varphi_{\nu-} = -\varphi_{\nu+} = \phi_{\nu}/2$ $(\Phi_{\nu}=0)$ and for the normalized exchange energies $V_{\mathrm{ex}}/\Delta=5$ [Fig.~\ref{fig-TrSgRatio}(a)], $V_{\mathrm{ex}}/\Delta=20$  [Fig.~\ref{fig-TrSgRatio}(b)]. Moreover, the figure displays the ratio $b_{\mathrm{tr}/\mathrm{sg}}$ for different phase differences $\phi_z$
and fixed phase difference $\phi_y=2\pi/3$. As can be seen from  Fig.~\ref{fig-TrSgRatio}(a), for small normalized exchange energies $V_{\mathrm{ex}}/\Delta$ the ratio $b_{\mathrm{tr}/\mathrm{sg}}$ is peaked where the peak position depends on the values of the superconducting phase differences $\phi_{\nu}$ and the resistance ratios $r_{\nu}$. Here, the ratio $b_{\mathrm{tr}/\mathrm{sg}}$ is smaller than one for the whole energy range $\varepsilon/\Delta \leq 1$. Interestingly, the situation changes drastically upon increasing the normalized exchange energy $V_{\mathrm{ex}}/\Delta$ as shown in Fig.~\ref{fig-TrSgRatio}(b). In this case, the ratio  $b_{\mathrm{tr}/\mathrm{sg}}$ may be significantly larger than one for small normalized energies $\varepsilon/\Delta$. This can be easily understood by
considering in Eq.~(\ref{eq-Def-MFns}) the limit $|M_{3N}-  iv_{\mathrm{ex}}|\gg f_{S} \sqrt{m_{2}^{2}+m_{1}^{2}}$, which corresponds to the limit $V_{\mathrm{ex}}/\Delta \gg 1$, so that $b_{\mathrm{tr}/\mathrm{sg}}=\vert\sin(\alpha)v_{\mathrm{ex}}/M_{3N}\vert \propto |\sin(\alpha) V_{\mathrm{ex}}/\varepsilon|$. Thus, we obtain in the limit $\varepsilon/\Delta \rightarrow 0$ for large normalized exchange energies a ratio $b_{\mathrm{tr}/\mathrm{sg}}$ that may be larger than
one.

In Fig.~\ref{fig-TrSgRatio}(b), where we have chosen  $V_{\mathrm{ex}}/\Delta=20$, the ratio $b_{\mathrm{tr/sg}}$ turns to zero for small normalized energies $\varepsilon/\Delta$ and the phase differences $\phi_z=0$ (solid line) and $\phi_z=\pi/2$ (dashed line) because in this case the normalized exchange energy $V_{\mathrm{ex}}/\Delta$ is still too small to shift the ratio $b_{\mathrm{tr/sg}}$ in this energy range to values  significantly larger than one. Effectively, we observe that the curves in Fig.~\ref{fig-TrSgRatio}(b) approximately converge to a single one with the energy dependence $b_{\mathrm{tr}/\mathrm{sg}} \propto |\sin(\alpha) V_{\mathrm{ex}}/\varepsilon|$ for really large normalized exchange energies $V_{\mathrm{ex}}/\Delta \gg 1$. Moreover, for large enough values of $V_{\mathrm{ex}}/\Delta$, the function $b_{\mathrm{tr}/\mathrm{sg}}$ is always larger than one for the relevant normalized energies $\varepsilon/\Delta < 1$, i.e., for large normalized exchange energies, the triplet component always dominates the singlet component.

Note that the DOS and the Josephson current are determined by the $\hat{\tau}_{3}\hat{\sigma}_{0}$ component ${\check{g}_{30}=(\mathrm{Re}[\mathcal{D}_{F+}^{-1}]M_{3N}+\mathrm{Im}[\mathcal{D}_{F+}^{-1}]v_{ex})\hat{\tau}_{3}\cdot\hat{\sigma}_{0}}$ and the singlet component $\check{f}_{\mathrm{sg}}$ of $\check{g}$, respectively. The latter is the case because in the commutator at the right-hand side of Eq.~(\ref{eq-BoundaryConds}) no contributions proportional to $\hat{\tau}_3 \cdot \hat{\sigma}_0$ emerge from $[\check{f}_{\mathrm{tr}},\check{g}_{S}(\zeta L_{\nu})]$. In particular, the Josephson current is determined by Eq.~(\ref{eq-supercurrent}) in which the factor $\mathcal{D}_N^{-1}$ should be replaced by $\mathrm{Re}[\mathcal{D}_{F+}^{-1}]$.

It is important to emphasize the difference between the considered case of short F-wires (one can speak of a ``zero"-dimensional cross) and the case of long F-wires. In the latter case the charge near the SF interface is carried by singlet and triplet Cooper pairs but at distances from the SF interface larger than $\xi_{F} \sim \sqrt{D_{F}/V_{ex}}$ the singlet and short range triplet components vanish and the charge current is only due to the so-called long-range triplet component (LRTC). Thus, a conversion of the singlet and short range triplet components into the LRTC takes place. In the case of short F-wires considered in this paper such a conversion does not occur and all the current components are constant in space.

Note that in cross geometry with long F-wires ($L_{y,z} \gg \xi _{F}$) and space-independent exchange energy in each single wire, the LRTC can not arise because the singlet and short range triplet components vanish near each SF interface on the distances of the order of $\xi_{F}$.

In the considered ``zero"-dimensional model, it is possible to calculate the singlet and triplet current components at the interfaces defined by
\begin{equation}
j_{\mathrm{sng/tr}} \sim \sum_{\omega} \mathrm{Tr} \left\{ \hat{\tau}_3 \cdot \hat{\sigma}_0 \cdot [\check{f}_{\mathrm{sng/tr}} \cdot \partial_\nu \check{f}_{\mathrm{sng/tr}}] \right\}\Big|_{\zeta L_{\nu}}.
\end{equation}
The functions $\check{f}_{\mathrm{sng/tr}}$ are given by Eqs.~(\ref{eq-fsng}-\ref{eq-ftr}). The derivatives $\partial_\nu\check{f}_{\mathrm{sng/tr}}|_{\zeta L_{\nu}}$ are easily obtained from the boundary conditions~(\ref{eq-BoundaryConds}) which can be rewritten in the form
\begin{equation}
\partial_{\nu}\check{g}\big|_{\nu=\zeta L_{\nu}}
=
\zeta \kappa_{\nu} \left( \check{g}_S - \check{g} \cdot \check{g}_S \cdot \check{g} \right)\big|_{\nu=\zeta L_{\nu}}
\end{equation}
by extracting terms proportional to $\hat{\tau}_1~\cdot~\hat{\sigma}_3$~and~$\hat{\tau}_2~\cdot~\hat{\sigma}_3$ (singlet component) and all the other terms non-diagonal in the particle-hole space (triplet component). The expressions for $\check{g}$ and $\check{g}_{S}(\zeta L_{\nu})$ are presented in Eqs.~(\ref{eq-gcheckfinal}) and (\ref{eq-gScheck}). The calculations can be shortened by choosing a rotated basis [cf. Eq.~(\ref{eq-TransformationByU})], i.e., using $\check{\mathfrak{g}}$ instead of $\check{g}$, since the current contributions as well as the matrices $\check{g}_{S}(\zeta L_{\nu})$ are invariant under such a rotation.

\begin{figure}[!t]
\includegraphics[scale=0.25]{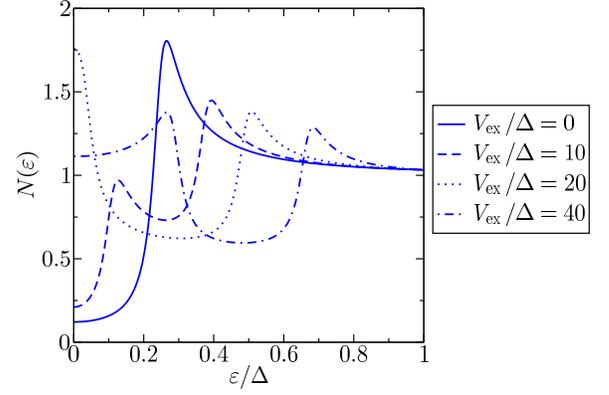}
\caption{(Color online) Density of states $N(\varepsilon)$ in the n$-$wire as a function of normalized energy $\varepsilon/\Delta$ for the case  $\varphi_{\nu+} = -\varphi_{\nu-} = \phi_{\nu}/2$ $(\Phi_{\nu}=0)$, $\phi_z = 2\pi$, $\phi_y = 2\pi/3$, and different exchange energies $V_{\mathrm{ex}}/\Delta$. The plots are presented for the parameters $\gamma/\Delta = 0.03$, $r_z = r_y = 1$, and $L_z/\xi_S = L_y/\xi_S = 0.1$.}
\label{fig-DOS-ferromagnet}
\end{figure}

The singlet current $j_{\mathrm{sng}}$ obtained in this way coincides with the total current obtained directly from the right-hand side of the boundary condition, whereas the triplet current $j_{\mathrm{tr}}$ vanishes. This shows that the current through the interfaces in the considered problem only consists of a singlet component. In Fig.~\ref{fig-DOS-ferromagnet} we show the DOS in the n$-$wire as a function of normalized energy $\varepsilon/\Delta$ for the case $\varphi_{\nu+} = -\varphi_{\nu-} = \phi_{\nu}/2$ $(\Phi_{\nu}=0)$
with $\phi_z = 2\pi$, $\phi_y = 2\pi/3$ and for different exchange energies $V_{\mathrm{ex}}/\Delta$.

We observe that the effect of the normalized exchange energy  $V_{\mathrm{ex}}/\Delta$ on the DOS is a splitting of the peak for vanishing $V_{\mathrm{ex}}/\Delta$, i.e., the gap induced in the n$-$wire is shifted to lower normalized energies. Increasing the normalized exchange energy $V_{\mathrm{ex}}/\Delta$ the two peaks initially move further apart from each other and their amplitudes decrease. If the peak located at lower normalized energy approaches the origin by increasing $V_{\mathrm{ex}}/\Delta$ furthermore, it bounces back from the vertical axis and also moves to higher energies together with the other peak. Thus, once the peak at lower  normalized energy has bounced back from the vertical axis, the induced gap in the n$-$wire vanishes and the DOS approaches $1/2$ between the two peaks and $1$ at the tails. Finally, as the normalized exchange energy is increased further the region where the DOS approaches $1/2$ becomes narrower and the two peaks move to higher energies.

This feature has also been obtained in Ref.~\onlinecite{Tanaka07} where a diffusive ferromagnet attached to a normal metal and a superconducting reservoir was considered. It is reported that for certain values of exchange field a peak at zero energy appears in the DOS which corresponds to the generation of the triplet pairing amplitude at low energy. The authors elaborated that the form of the DOS in the ferromagnet for low energies can be used to find the pairing symmetry in the ferromagnet, i.e., whether the peak is generated by the singlet or triplet pairing states. From this analysis, we can conclude that since in Fig.~\ref{fig-TrSgRatio}(b) for $\phi_z = 2\pi$ (dash-dotted line) the triplet component dominates the singlet component the peak at zero energy in Fig.~\ref{fig-DOS-ferromagnet} for $V_{\mathrm{ex}}/\Delta = 20$ (dotted line) is generated by the triplet component at low energies.

Note that the influence of the triplet component on the DOS was also analyzed in Refs.~\onlinecite{Nazarov,Volkov03,Volkov05} for other systems. Also, the splitting of the main peak in the DOS due to the exchange field was studied earlier in Refs.~\onlinecite{Golubov02,Koshina02,Tanaka06.1,Tanaka06.2}.

\begin{figure}[!t]
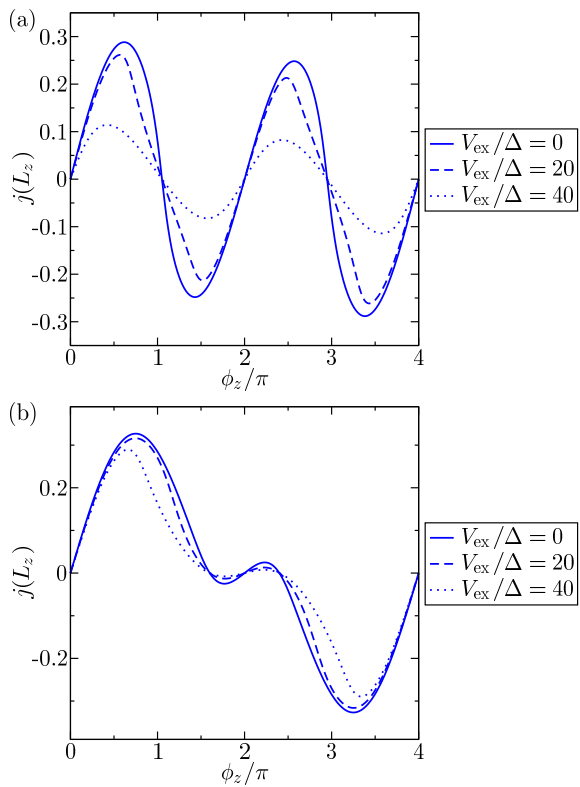

\includegraphics[scale=0.25]{Fig8a-Jz_F1_rz1-ry0.1.eps}\vspace{1mm}
\includegraphics[scale=0.25]{Fig8b-Jz_F1_rz1-ry1}
\caption{(Color online) Normalized supercurrent $j(L_z)$ in the horizontal wire at the right interface as a function of the horizontal phase difference $\phi_z$ for different values of the normalized exchange energy  $V_{\mathrm{ex}}/\Delta$. The plots are presented for the case $\varphi_{\nu+} = -\varphi_{\nu-} = \phi_{\nu}/2$ $(\Phi_{\nu}=0)$ with $\nu=z,y$ and the parameters $T/\Delta = 0.05$, $r_z = 1$, $\phi_y = 2\pi/5$, $L_z/\xi_S = L_y/\xi_S = 0.1$, (a) $r_y = 0.1$, and (b) $r_y = 1$.}
\label{fig-jz-case1-ferromagnet1}
\end{figure}
In order to figure out the influence of the normalized exchange energy $V_{\mathrm{ex}}/\Delta$ on the supercurrent, in  Fig.~\ref{fig-jz-case1-ferromagnet1}, we plot the normalized Josephson current at the right horizontal interface, i.e., at $z=L_z$, for the case $\varphi_{\nu+} = -\varphi_{\nu-} = \phi_{\nu}/2$ $(\Phi_{\nu}=0)$ as a function of the phase difference $\phi_z$ with the parameters $\phi_y = 2\pi/5$ $L_z/\xi_S = L_y/\xi_S = 0.1$, $r_z = 1$, $r_y = 0.1$ [see Fig.~\ref{fig-jz-case1-ferromagnet1}(a)], $r_y = 1$  [see Fig.~\ref{fig-jz-case1-ferromagnet1}(b)], and for different values of the exchange energy $V_{\mathrm{ex}}/\Delta$.

Here, we find that the effect of the normalized exchange energy  $V_{\mathrm{ex}}/\Delta$ on the supercurrent is a distortion of the  sinusoidal shape while the $2\pi$ periodicity for $r_z = 1$, $r_y = 0.1$ is conserved [see Fig.~\ref{fig-jz-case1-ferromagnet1}(a)]. With increasing  normalized exchange energy the supercurrent is more and more distorted leftwards together with a decrease in amplitude, i.e., the exchange energy can significantly reduce the magnitude of the supercurrent. For the case $r_z = r_y = 1$, Fig.~\ref{fig-jz-case1-ferromagnet1}(b), the effect of the  normalized exchange energy is weaker compared to the case $r_z = 1$, $r_y = 0.1$, Fig.~\ref{fig-jz-case1-ferromagnet1}(a). Here, much higher values of $V_{\mathrm{ex}}/\Delta$ are necessary in order to reduce the magnitude of $j(L_z)$ considerably but the local extremes close to $\phi_z = 2\pi$ are smoothened more easily. For the case $r_z \ll r_y$ the form of the  supercurrent is nearly unchanged by the normalized exchange energy $V_{\mathrm{ex}}/\Delta$ and retains its sinusoidal behavior with a $4\pi$
periodicity [see Fig.~\ref{fig-JosephsonCurrent-rY}(a)].

Comparing the critical Josephson current $j_c(L_z)$ at the right horizontal interface with respect to the normal metal wire and the ferromagnetic wire, we find that the magnitude of $j_c(L_z)$ decreases with increasing normalized exchange energy $V_{\mathrm{ex}}/\Delta$, while the form of the critical Josephson current remains unchanged (see   Fig.~\ref{fig-CriticalJosephsonCurrent}). For a ferromagnetic wire the decreasing effect of the exchange energy is barely visible for $r_z \gg r_y$ and $r_z \ll r_y$ while it is slightly stronger for $r_z = r_y$.

\section{Conclusion}

We have calculated the Josephson dc current in a multiterminal structure of cross-type geometry, which consists of four superconductors and  one-dimensional normal or ferromagnetic wires connecting the superconducting electrodes. The length of the wires is assumed to be short, i.e., the Thouless energy exceeds any characteristic energy of the system ($\Delta $, $T$, or $V_{\mathrm{ex}}$). In this case, one can easily obtain a solution of the Usadel equation for the quasiclassical matrix Green's functions $\check{g}$ in a compact form. With the help of the matrix $\check{g}$ we find the DOS and the dc Josephson current $I_J$. It turns out that both the DOS and $I_J$ strongly depend on the phase differences between opposite superconductors. The sinusoidal dependence of $I_J$ on the phase difference $\phi_z$ has different period which is determined by the ratio of the interface resistances $R_{Sn,z}/R_{Sn,y}$. If this ratio is small, the period is equal to $2\pi$, in the opposite limit the period equals $4\pi$. It is found that the DOS in the wire has a minigap that depends on the phase difference $\phi_{\nu}$ and can turn to zero.

We have obtained also simple formulas for the singlet and triplet components induced in the ferromagnetic wires. The odd-frequency $s$-wave triplet component has non-zero projections of the total spin $S$ on the directions on the wires and can be larger than the amplitude of the singlet component. However, the Josephson current is caused only by the singlet component. The obtained results may be useful in superconducting electronics and spintronics (see Ref.~\onlinecite{Linder12} and references therein).

\end{document}